\documentclass[aps,pre,twocolumn,floatfix]{revtex4}
\usepackage{graphicx,color}
\usepackage{amsmath,amssymb}
\begin{document}

%%%%%%%%%%%%%%%%%%%%
%
\title{When a DNA Triple helix melts: An analog of the Efimov state}

\author{Jaya Maji} 
\email{jayamaji@iopb.res.in}
\author{Somendra M. Bhattacharjee}
\email{somen@iopb.res.in}
\affiliation{ Institute of Physics, Bhubaneswar 751005, India}
\author{Flavio Seno} 
\email{flavio.seno@pd.infn.it}
\author{Antonio Trovato}
\email{antonio.trovato@pd.infn.it}
\affiliation{CNISM, Dipartimento di Fisica, Universit\`a di
             Padova, Via Marzolo 8, 35131 Padova, Italy} 
%
%%%%%%%%%%%%%%%%%%%%FIGURES%%%%%%%%%%%%%

\newcommand{\figgaus}{%
\begin{figure}[htbp]
   \centering
   \includegraphics[width=0.45\textwidth]{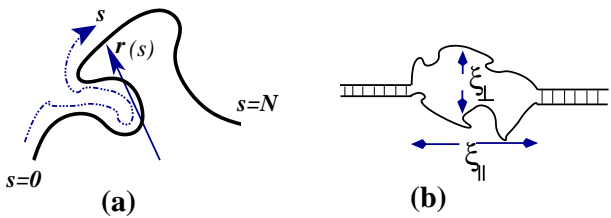}
%%%%%% fig1.eps goes here %%%%%%%%%%%%%

   \caption {(a) A Gaussian chain or a random walker in three
     dimensions. An arbitrary point is specified by the contour length
     $s$ and its position vector ${\bf r}(s)$, with the two ends at
     $s=0$ and $s=N$.  (b) A bubble on a two chain system.  The extent
     along the contour is $\xi_{\parallel}$ and the spatial extent is
     $\xi_{\perp}$. The ladder-like regions represent the bound states
     with base-pairing between points with same $s$.  }
   \label{fig:gaus}
 \end{figure}
}
%%%%%%%%%%%%%%%%%%%%%

%%%%%%%%%%%%%%%%%%%%%%
\newcommand{\figvolley}{%
\begin{figure}[htbp]
   \centering
   \includegraphics[width=0.45\textwidth]{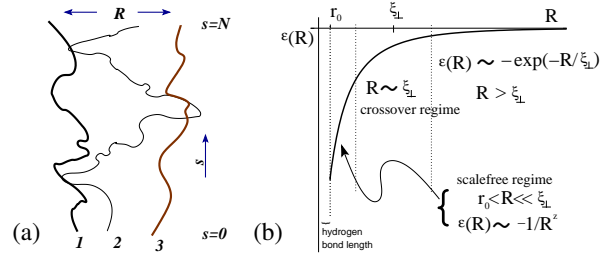}
%%%%%% fig2.eps goes here %%%%%%%%%%%%%%%%%%%%

   \caption {(a) Two noninteracting Gaussian chains 1 and 3 separated
     by a distance $R$.  Each of these can pair with a flexible chain
     2 denoted by a thin line.  The polymers are shown with $s$ as an
     axis. In the quantum correspondence, these polymers are the paths
     with $s$ as the time like axis. (b) The effective interaction
     $\epsilon(R)$ between 1 and 3. Triple chain bound state (Efimov
     effect) occurs in the region $r_0<R<<\xi_{\perp}$ and extends
     over the whole range for $\xi_{\perp}\to \infty$.  }
   \label{fig:vol}
 \end{figure}
}
%%%%%%%%%%%%%%%%%%%%%

%%%%%%%%%%%%%%%%%%%%%%%%%%%%%%%%%%%%%%%%%%%%%%%

\newcommand{\fighier}{%
\begin{figure}[htbp]
   \centering
   \includegraphics[width=0.35\textwidth]{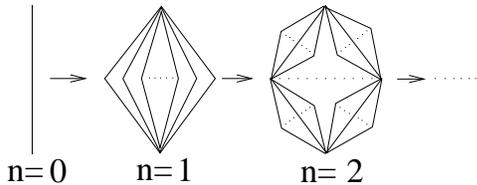}
%%%%%%%% fig3.eps goes here %%%%%%%%%%%%%%%%%%

   \caption{Recursive construction of the hierarchical lattice used
     for the real space renormalization group.  At every stage, each
     line is replaced by a diamond of $2b$ lines.  }
   \label{fig:hier}
 \end{figure}
}
%%%%%%%%%%%%%%%%%%%%%

\newcommand{\figflow}{%
\begin{figure}[htbp]
   \centering
   \includegraphics[width=0.45\textwidth]{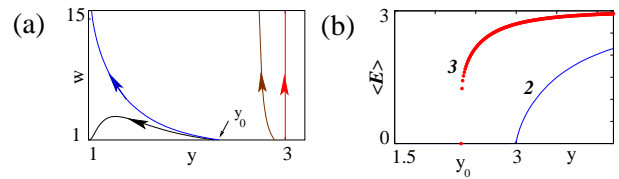}
%%%%%%%%%%%%%% fig4.eps goes here %%%%%%%%%%%%%%%%%%%

   \caption{(a)RG Flow-diagram in the $y$-$w$ plane for the symmetric
     case $y_{12}=y_{23}=y_{13}$, all starting with $w=1$. Here $b=4$.
     The flow of $w$ goes to $\infty$ if the starting
     $y>y_0=2.32402...$, otherwise to $1$ (high temperature fixed
     point).  The trajectories with starting $y<y_0$ end at $w=y=1$.
     (b) Average energy per monomer vs.  temperature from direct
     computation (chain length=$2^{25}$). For two chains (marked 2)
     average energy undergoes a continuous transition at $y=y_c$ while
     the average energy for three chains (marked 3) shows a jump at
     $y=y_0$.  The region from $y_0$ to $y_c$ is the Efimov-like three
     chain bound state.  }
   \label{fig:1}
 \end{figure}
}
%%%%%%%%%%%%%%%%%%%%%

\newcommand{\figcontour}{%
\begin{figure}[htbp]
   \centering
   \includegraphics[width=0.35\textwidth]{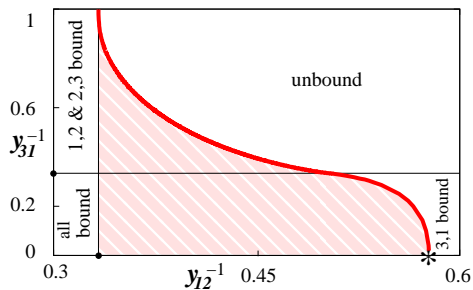}
%%%%%%%%%%%%%%%%%%%% fig5.eps goes here %%%%%%%%%%%%%%%%%%

   \caption{ Phase Diagram: $y_{31}^{-1}$ vs. $y_{12}^{-1}$
     ($y_{12}=y_{23}$), for $b=4$. The duplex melting point at
     $y_c^{\{ij\}}=b-1$ is indicated by the horizontal and vertical
     lines.  Three chains are bound in the shaded region with the
     thick curve representing the three chain bound-unbound
     transition.  Above the horizontal line at $y_{31}=b-1$ in the
     shaded region, a triplex state exists even though no two should be
     bound.  The bound states in other regions are as indicated. The
     star at $y_{12}^{-1}=1/\sqrt{b-1}$ is the melting of chain 2 and
     composite 1,3.  }
   \label{fig:2}
 \end{figure}
}
%%%%%%%%%%%%%%%%%%%%%%%%

\newcommand{\figmodel}{%
\begin{figure}[htbp]
   \centering
  \includegraphics[width=0.2\textwidth,angle=90]{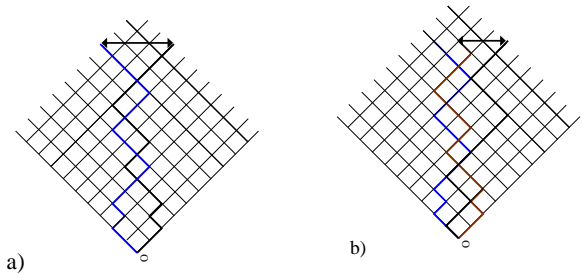}
%%%%%%%%%%%%%%%%%% fig6.eps goes here  %%%%%%%%%%%%%%%%%%%%%%%

   \caption{a) Graphical representation of directed two strand DNA in
     1+1 dimensions.  The two strands start from the same origin and
     are directed along the direction (1,1). The walks can cross each
     other. Each bubble opening is weighted with a factor $\sigma$, each
     interaction (overlap of the two chains) with a weight $y$. The
     weight of this conformation is therefore $\sigma^5 y^5$. Notice
     that each interaction corresponds to sites with the same index.
     b) The three chain models. A factor $\sigma$ is given to each
     bubble opening among all possible pairs of polymers. For model A the
     weight of this configuration is $ \sigma^{11} y^{12}$ (see text) for
     model B, where the interaction is only between the red and the
     blue chain, and the red and the brown chain, the weight is
     $\sigma^{11} y^9 $.  The black arrows, in both figures, indicate the
     end to end distance $r_N$ used to estimate the melting
     transition.}

   \label{fig:3}
 \end{figure}
}
%%%%%%%%%%%%%%%%%%%%%%%%

\newcommand{\figsimulation}{%
\begin{figure}[htbp]
   \centering
   \includegraphics[angle=270,width=7cm,clip]{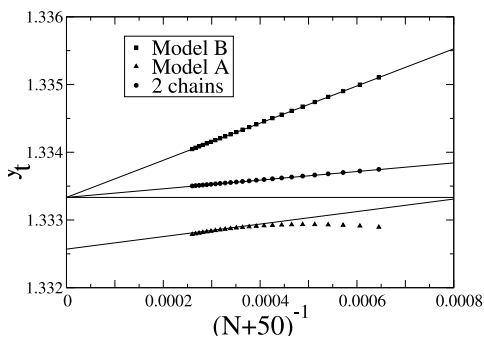}
%%%%%%%%%%%%%%%%% fig7.eps goes here %%%%%%%%%%%%%%%%%%

   \caption{ For models A and B described in the text and for a two
     strand model, we estimate the denaturation transition by looking,
     in the three cases, at the crossing $y_t$ of the curves
     $\xi_N(y)$ and $\xi_{N+100}$, where $N$ is ranging from $1500$ to
     $3900$ in steps of 100. These crossing points are plotted as a
     function of $1/({N+50})$.  The case $\sigma=0.5$ is considered.
     The horizontal continuous line represents the exact melting value
     $y_t=4/3$ of the duplex model. Linear extrapolation curves are
     also shown. Their intercepts with the y-axis are $1.3333\pm
     0.0001$ for model B, $1.3333 \pm 0.0001$ for the 2 chains, and
     $1.3326 \pm 0.0001$ for model A.  These results definitely show
     that model B has the same melting transition as the duplex model,
     whereas model A melts at a higher temperature (Efimov effect) }
    \label{fig:4}
 \end{figure}
}
%%%%%%%%%%%%%%%%%%%%%%%%%%%%%%%%%%%%%%%%
%%%%%%%%%%%%%%%%%%%%%%%%%%%%%%%%%%%%%%%%%%%%%%%

%%%%%%%%%%%%%%%%%%%%%%%%%%%%%%%%%%%%%%%%%%%%%%%

\newcommand{\figorder}{%
\begin{figure*}[htbp]
   \centering
   \includegraphics[width=6.in]{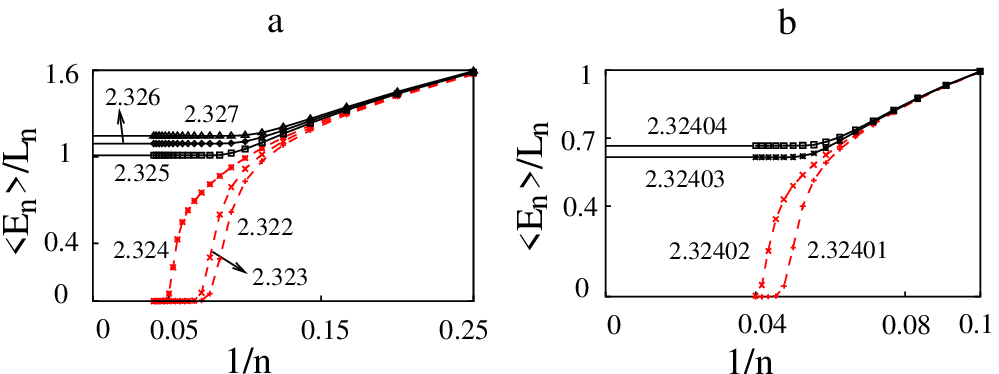}
%%%%%%%% energXXX.eps goes here %%%%%%%%%%%%%%%%%%

   \caption{Plots of $E_n/L_n$ vs $1/n$, $E_n$ being the three chain
     average energy, $n$ the generation number. We went upto $n=25$
     for which the length of each polymer is $L_n=2^{25}$.  In (a) we
     show for $y=2.322+0.001*n, n=$0-5, while a finer grid result is
     shown in (b) with $y=2.32400+0.00001*n, n=$1-4.  The lines
     show the extrapolations to $n\to\infty$.  The discontinuity at
     the transition is visible.  }
   \label{fig:ord}
 \end{figure*}
}
%%%%%%%%%%%%%%%%%%%%%

% 

\begin{abstract} 

  The base sequences of DNA contain the genetic code and to decode it
  a double helical DNA has to be unzipped to reveal the bases.  Recent
  studies showed that a third strand can be used to identify the base
  sequences, not by opening the double helix but rather by forming a
  triple helix.  It is predicted here that a three stranded DNA
  exhibits the unusual behaviour of the existence of a three chain
  bound state in the absence of any two being bound.  Such a state can
  occur at or above the duplex melting point.  This phenomenon is
  analogous to the Efimov state in three particle quantum mechanics.
  A scaling theory is used to justify the Efimov connection.  Real
  space renormalization group, and exact numerical calculations are
  used to validate the prediction of a biological Efimov effect.
\end{abstract}

\date{\today}  

\maketitle

\section{Introduction}\label{sec:intro}
Double helical DNA is common, but under certain circumstances DNA can
form a triple helix too.  The 1957-discovery of a 3 strand DNA
remained a curiouser till the recognition in 1987 that a third strand
DNA can actually recognize the base sequence of the double helix even
without opening it\cite{rich,helen-1,helen-2}.  Owing to its enhanced
stability that can affect activities like gene expression, DNA
replication and others requiring DNA opening, triple helix kindled new
hopes in therapeutic applications\cite{jain}.  Till date it has been
possible to make and study triple helices in vitro, amidst high hopes
of their relevance in vivo\cite{triple-1,triple-2}.  At ambient
temperatures, the double helix is formed with classical Watson-Crick
base pairing while the third strand forms non-classical Hoogsteen or
(reverse Hoogsteen) base pairing with one of the other two.  Triple
helix can also be formed with DNA-RNA \cite{rna} and DNA-peptide
nucleic acid (PNA) whose uncharged peptide backbone helps in
stabilization of the triplet structure\cite{pna-1,pna-2}.  On a
completely different front, in 1970 Efimov in his studies on nucleons,
showed an effect, now bearing his name, in 3-body nonrelativistic
quantum mechanics, viz., the possibility of a three body bound state
where none of the pairs are bound and the overall size of the bound
state is much larger than the range of the pair
potentials\cite{efi-1,efi-2,efi-3,fonseca,efirev}.  One purpose of
this paper is to wed these two disparate systems to show that an
analog of the quantum Efimov state is the triple helix {\it at or
  above} the double helix melting point.

The origin of the Efimov effect is the scale-free quantum fluctuation
near the zero-energy threshold of two body binding.  The effective
interaction that gives the three-body binding does not depend on the
detailed nature of the short-range interactions and is operative
outside its range. After many years since its discovery, it is now
seen in systems over various length-scales ranging from nucleons (halo
nucleus) to atomic physics and ultra-cold atoms under Feshbach
resonance\cite{efiexp-1,efiexp-2,efiexp-3}.  A triple stranded DNA
near its melting point is shown to be a unique example from the domain
of classical biology and might provide an affordable system for
studying aspects of the quantum Efimov physics.

There have been many physical, chemical and biological studies of
triplex forming oligonucleotides (TFO) with applications in
mind\cite{plum}.  We take a thermodynamic point of view where the long
chain limit is taken to explore the nature of phases and transitions
of a triplex.  The case of oligonucleotides can then be formulated by
studying the finite-length effects on the transitions of the infinite
chain system.  Our focus in this paper is mainly on the infinite chain
limit.  A scaling approach is used in this paper to see the origin of
the Efimov effect via the development of a ubiquitous attractive
$1/r^2$ interaction over a range beyond the short range of the
pair-potential.  The short range in the context of DNA is the hydrogen
bond length or the base pair separation.  The Efimov interaction
requires a pairwise attraction at the critical threshold and therefore
a quantum system in $d\leq 2$ will not show this because in these
lower dimensions a bound state exists for any shallow potential.
However in a polymeric system if a finite melting point can be induced
then the Efimov effect would be visible.  In subsequent sections, we
consider a few simplified polymer models to establish the Efimov
effect in DNA.  A renormalization group (RG) approach is used to show
the Efimov-like three chain phase in $d\geq 2$ models and a high
precision numerical approach is used for $1+1$ dimensional models. The
phase diagram for the bubble-free three chain fork model, with a first
order duplex melting, is discussed in the Appendix A.  Some details of
the RG calculation can be found in Appendix B.

\section{Scaling approach to the Efimov effect}\label{sec:efi}
Let us consider three Gaussian polymers interacting with one another
through a DNA base pairing type short range interaction.  A monomer of
a chain $j$ ($j=1,2,3$) is identified by a length variable $s$
measured along the contour of the chain with ${\bf r}_j(s)$ as its
position vector.  See Fig.~\ref{fig:gaus}a.  This (and the models in
the subsequent sections) are coarse-grained models where what we call
a monomer in fact represents several base-pairs.  The interaction
involves monomers with same $s$ of different chains, so that the
Hamiltonian for strands of length $N$ ($N\to\infty$) can be written in
a dimensionless form as\cite{smbunz}
\begin{equation}
  \label{eq:1}
  \beta H=  \int_0^N\!  ds\left[ \sum_{j=1}^{3} \frac{K_j}{2}
  \left(\frac{\partial {\bf r}_j(s)}{\partial s}\right )^2
+\sum_{k<l}  V_{kl}({\bf r}_k(s),{\bf r}_l(s)) \right ],
\end{equation}
where $V_{kl}$ is the short-range attractive interaction representing
the base pairing between chains $k$ and $l$, and $K_j$ is the elastic
constant or flexibility of chain $j$, $\beta=1/k_BT$, with $T$ as
temperature, $k_B$ the Boltzmann constant.  The first term on the rhs
of Eq.~(\ref{eq:1}) represent the elastic energy or the connectivity
of the polymers.  The partition function is given by
\begin{equation}
  \label{eq:2}
  Z=\int {\cal DR} \exp(-\beta H),
\end{equation}
where $\int {\cal DR}$ stands for the summation over all
configurations or paths of the three chains with appropriate boundary
conditions.  We may choose ${\bf{r}}_j(0)=0$ for all $j$, i.e., the
three chains are tied at the origin at one end.  The other ends may be
free.

\figgaus

By treating the partition function as a path integral, the imaginary
time transformation $s=it$ converts $Z$ to a propagator in quantum
mechanics for three particles with a pairwise interaction $V_{kl}({\bf
  r}_k,{\bf r}_l)$, and the masses determined by $K_j$.  With a short
range attractive potential of range $r_0$, there is a critical
threshold of the potential in three dimensions for two particles to
form a bound state, and for stronger potentials, there will be a
finite number of bound states.  At the critical coupling the two
particle system has a zero-energy bound state with infinite width of
the eigenfunction or infinite scattering
length\cite{efi-1,efi-2,efi-3}.

\figvolley

The DNA-quantum correspondence relates the quantum critical threshold
to the thermal melting of duplex DNA, a continuous transition in this
model, with a diverging length scale\cite{Fisher,gotoh}.  This scale
can be associated with the transient bubbles that may form for
temperatures just above the melting point $T_c$ (analogous to the
scattering length $a$ of the quantum problem) or the fluctuation in
size and shape of the bubbles below $T_c$ (width of the wave function
in the quantum case).  The bubbles can be characterized by two scales,
$\xi_{\perp}$ for the spatial extent and
\begin{equation}
  \label{eq:3}
\xi_{\parallel}\sim \xi_{\perp}^z,  
\end{equation}
for the length of the bubble (Fig.~\ref{fig:gaus}b) where $1/z$ is like
a size exponent for polymers.  For the quantum problem, with
$\xi_{\perp}$ as the scattering length $a$, $z$ corresponds to the
dynamic exponent.  The diffusive or Gaussian nature of the free chain
for our case (or the quantum problem) implies
\begin{equation}
  \label{eq:7}
z=2.  
\end{equation}
One can see this value of $z$ from the scale invariance of the elastic
energy term in Eq.~(\ref{eq:1}).  This term ensures the connectivity
of a polymer and remains invariant under a scale transformation, ${\bf
  r}\to\lambda {\bf r}, s\to \lambda^z s$, with Eq.~(\ref{eq:7}).  A
similar invariance argument yields Eq.~(\ref{eq:7}) for the
nonrelativistic free particle Schr\"odinger equation.
 
Suppose we have two noninteracting chains 1 and 3, both interacting
with another one, chain 2.  For simplicity, though not essential, we
may take 1 and 3 as relatively stiffer compared to 2, and they are at
a distance $R$ apart (see Fig.~\ref{fig:vol}).  Choose the pairwise
interaction between 1-2 and 2-3 or the temperature close to the
critical threshold.  In that case, if $\xi_{\perp} >R$, then inbetween
two contacts with chain 1, chain 2 is expected to meet chain 3.  The
interaction is attractive.  This exchange over a large length scale
induces an attraction between 1 and 3 if chain 2 is integrated out
from the partition function of Eq.~(\ref{eq:2}).  The effective
interaction $\varepsilon(R)$ is the change in free energy $\Delta F/N$
because of the presence of a scale $R$ so that it can be written as
\begin{equation}
  \label{eq:5}
  \Delta F\sim - \frac{N}{\xi_{\parallel}} \; 
    {\cal F}\left({R}/{\xi_{\perp}}\right), 
\end{equation}
where the first factor is the number of blobs and ${\cal F}(x)$ is a
scaling function.  For $\xi_{\perp}\to\infty$, Eq.~(\ref{eq:5}) should
be independent of it requiring ${\cal F}(x) \sim x^{-z}$, as ${x\to
  0}$, so that in this limit, by Eq.~(\ref{eq:7}),
\begin{equation}
  \label{eq:6}
  \varepsilon(R)\equiv \frac{\Delta F}{N} =-\frac{A}{R^{z}} =
  -\frac{A}{R^{2}}, \ ( A={\rm a\  constant}). 
\end{equation}
We see the emergence of a ``universal'' $1/R^2$ interaction for a
region $r_0<R\ll\xi_{\perp},$ and this attractive long range
interaction can produce a bound-state of size bigger than $r_0$.  To
be noted here, that above the duplex melting point, there will be
large bubbles, thanks to critical fluctuations, and so the attractive
long range interaction persists even above the melting point.  This
gives the possibility of a three chain bound state where none are
bound in the two particle potential well.

In the quantum language the large fluctuation is the resonance, and
the Efimov effect is due to the attraction produced by the multiple
scattering of a light particle off the two heavier ones.  This is one
aspect of the Efimov effect.  In an exact study using a separable
potential (under the Born-Oppenheimer approximation) the effective
interaction between 1 and 3 by the hopping of 2 has been
calculated\cite{fonseca}.  The result of Ref. \cite{fonseca} can be
recast in our scaling language (note that  $\xi_{\perp}$ is the
scattering length $a$) as
\begin{equation}
  \label{eq:8}
  \varepsilon(R)= - \frac{1}{\xi_{\perp}^2} \; \frac{1}{\tilde{R}^{^z}} {\sf
    f}(\tilde{R}),\qquad {\rm where}\  \tilde{R}=\frac{R}{\xi_{\perp}},
\end{equation}
and 
\begin{equation}
  \label{eq:9}
  {\sf  f}(\tilde{R})= e^{-\tilde{R}} ( 2 \tilde{R}+  e^{-\tilde{R}}),
\end{equation}%
corroborating the scaling hypothesis of Eq.~(\ref{eq:5}).  One sees a
cross-over from $-1/R^2$ for $ \tilde{R}\ll 1$ to the Yuakawa form $-
e^{-\tilde{R}}/\tilde{R}$ for $ \tilde{R}\sim O(1)$.  The scale free
interaction permeates the whole region for $\xi_{\perp}\to\infty$ at
the critical or threshold value of the pair interaction.
Fig.~\ref{fig:vol}b shows the nature of the effective interaction in
presence of a scale $\xi_{\perp}$.  The Efimov interaction is beyond
the hydrogen bond length of duplex DNA.

The Efimov interaction allows a three chain bound state close to, or
just above the duplex melting point. The size of the bound state will
necessarily be much larger than the hydrogen bond strength.  The fact
that one sees Efimov states, though finite in number, for large but
finite $\xi_{\perp}$ (Eq.~(\ref{eq:9})), indicates that the triple
helix DNA may also show an Efimov analog bound state even above the
duplex melting temperature. This also means that the melting
temperature is higher for a triplex than for a duplex.  The minimal
model in Eq.~(\ref{eq:1}) has a continuous transition.  If other
interactions not in Eq.~(\ref{eq:1}) drive the transition first order,
the length scale $\xi_{\perp}$ will be non-diverging.  However, if the
transition is weakly first order, $\xi_{\perp}$ may be large enough to
accommodate the intermediate scale-free region shown in
Fig.~\ref{fig:vol}b allowing a bound state.  Hence in a weak first
order transition with large $\xi_{\perp}$ one would still see the
Efimov-like bound state.  A case of a strong first-order transition is
dealt with in Appendix A where it is shown that the effect is absent
if the bubbles are fully suppressed.

The second aspect of the Efimov effect is that at the critical point,
there is an infinite sequence of bound states in a geometric form
$E_n=E_0 (e^{-2\pi/s_0})^n$ where $s_0$ is system specific number
(e.g. for 3 identical bosons, $s_0=1.00624$).  The energy scale $E_0$
is also system-parameter dependent.  The above analysis is done with
the ground-state dominance assumption that is justifiable for
$N\to\infty$.  The polymer partition function of Eq.~(\ref{eq:2}) for
finite $N$ is given by the Efimov energies as
\begin{equation}
  \label{eq:10}
  Z\sim C_1 \exp(-E_0 N) + C_2 \exp(-E_1 N)+ ..., 
\end{equation}
where the temperature has been absorbed in the ``energies''.  The
terms beyond the ground state can be ignored if
\begin{equation}
  \label{eq:11}
N\gg 1/|E_1-E_0|\sim \frac{1}{\mid E_0 (1- e^{-2\pi/s_0})\mid}.  
\end{equation}
This is the length requirement for the triple DNA to show the
Efimov-like state.

The prediction of a large girth triplex close to (and even above) the
duplex DNA melting point awaits experimental tests.

\section {Efimov-like phase in   $d\geq 2$ }\label{sec:hier}
In order to justify the prediction of the scaling theory in a
systematic way, we adopt a renormalization group\cite{smsmb} approach
for $d>2$.  In the RG approach, the effects of interaction is probed
by summing over the configurations at a smaller scale (in the
partition function) and redefining the effective interaction on a
larger scale.  For a bound state, we should see an effective
interaction among the chains, irrespective of the scale of
coarse-graining. In contrast, for an unbound state, locally bound
monomers lose their importance as we sum over configurations and
therefore the effective interaction would vanish as the probing
lengthscale increases.  These effects are expressed by the RG flow
equations or recursion relations, as flows of the interactions with
length scale.  A two body bound state should therefore be possible if
the two body interaction does not vanish.  In the same spirit, a three
body bound state would occur if a three-body interaction becomes
operative, even if there is none to start with.  We express these RG
relations in an exact form on specially constructed hierarchical
lattices with discrete scaling symmetry and tunable dimensionality.

\fighier

Let us consider three directed polymers, which are doing random walk
from bottom to top on a hierarchical lattice, constructed recursively
with a motif of $2b$ bonds, as shown in Fig.~\ref{fig:hier}.  The
branching factor $b$ determines the effective dimension$(d)$ of the
hierarchical lattice as $d=\frac{\ln2b}{\ln2}$.  There are attractive
potentials $-\epsilon_{ij}$ and
$-\epsilon_{ijk}(\epsilon_{ij},\epsilon_{ijk}>0)$ if a single bond is
shared by two and three polymers respectively.  Although
$\epsilon_{123}=0$,  still it will be needed for the RG
transformation to probe the three-body bound state and it is generated
by renormalization.

The configurations of the two chain system can be classified as two
independent chains or inherently two chain configurations.  The
corresponding recursion relation for the two chain Boltzmann factors
$y_{ij}=\exp(\beta \epsilon_{ij})$ is given by\cite{smsmb}
\begin{equation}
y_{ij}^{'}=\frac{b-1+ y_{ij}^2}{b}.
\end{equation}

Similarly the three chain configurations 
can be classified as (i) three independent
chains, (ii) a combination of a double and a single chain, or (iii)
inherently three chain configurations, i.e. three chains sharing the
same bond.  By a decimation of the $2b$-bond motif, the recursion
relation for the three chain Boltzmann factor
$w=\exp(\beta\epsilon_{123})$ is given by
\begin{equation}
  w^{'}=\frac{(b-1)(b-2)+ (b-1){\displaystyle\sum_{i< j}}y_{ij}^2+
    w^2{\displaystyle\prod_{i< j}}{y_{ij}}^2}{{b^2}{\displaystyle\prod_{i<
        j}}{y_{ij}^{'}}},\ \ \ 
\label{eq:w}
\end{equation}
where $y', w'$ denote the renormalized values.

In the following discussion, we choose the initial value $w=1$.  The
three fixed points of $y_{ij}$ correspond to (1) $y^*=\infty$ (zero
temperature, pure bound state), (2) $y^*=1$ (infinite temperature, no
two body interaction), and (3) $y^*=(b-1)$ (two chain unstable
critical point).

In case there is no pairwise bound state or no pair interaction, $w$
has two fixed points $1$ and $b^2-1$.  The stable fixed point $1$
describes the high temperature fixed point or absence of three body
interaction and the unstable fixed point $b^2-1$ describes the
critical state produced by a pure three body interaction.  The flow
going to infinity is indicative of the three-chain bound state, formed
by the three body force, a case not of interest here.

\figflow

In case all pairs are in the critical state so far as the two body
interaction is concerned ($y_{ij}^*=b-1$), there is no real fixed
point for $w$.  The renormalization flow takes $w$ to infinity, as
shown in Fig.~\ref{fig:1}a. The three chains then form a bound state
at the two-body critical point.  For temperatures above the duplex
melting, {\it i.e.}, with initial values $y=y_{12}=y_{23}=y_{31}
<b-1$, the triplex will be in the denatured state if the flow goes to
$y=1,w=1$.  E.g., for $b=4$, the three chain melting is at
$y=2.32402...$ which is less than the duplex melting point $y_c=3$.  A
further confirmation of this triplex melting comes from an exact
numerical calculation of the average energy by iterating the partition
functions and their derivatives for large lattices.  This is shown in
Fig.~\ref{fig:1}b.  With $y\equiv y_{12}=y_{23}=y_{31}$, as in
Fig.~\ref{fig:1}a, the two strand system melts through a second order 
transition at $y=y_c$ (energy going continuously to $0$) whereas the
three strand system undergoes a first order transition at a
temperature $y=y_0<y_c$ (energy showing a discontinuity - see Appendix
B for details).  The region between $y=y_0$ to $y=y_c$, is the region
of a triple strand bound state when there should not be any duplex.

The phase diagram in the plane of $y_{13}^{-1}$ vs $y_{12}^{-1}$ with
$y_{12}=y_{23}, w=1$ is shown in Fig.~\ref{fig:2}.  For
$y_{13}^{-1}=0$, chains 1 and 3 are bound for ever and therefore chain
2 melts off at $y_{12}=\sqrt{b-1}$.  This point is indicated by a star
in Fig.~\ref{fig:2}.  Within the triangular shaded region bounded by
$y_{13}^{-1}=1/(b-1)$, $y_{12}^{-1}=1/(b-1)$, and the curved line
separating the unbound state, we have a triplex phase without pairing
of any two - the desired Efimov effect.

\figcontour

\section{Numerical simulation in $d=1+1$}
After arguing for the existence of an Efimov bound state for triple
DNA helices through the DNA- quantum correspondence and the scaling
argument, and after observing the effect on a hierarchical diamond
lattice, in this section we produce a clear numerical evidence that
the effect is present in an euclidean lattice even in $1+1$
dimensions.

Directed polymers in such a dimensionality are amenable to exact
solutions in the thermodynamic limit of infinite chain length or for
extremely accurate numerical simulations since their interchain
contacts are guaranteed to occur between monomers (see fig.
\ref{fig:3}) with the same index, as for DNA. For such reasons, they
played an important role in clarifying melting, cold unzipping and
overstretching properties of duplex DNA
\cite{ymodel,ymodel-2,ymodel-3}, and for the same reasons they turn
out to be the most convenient models in which to test numerically the
existence of the Efimov effect that on higher dimensions might be
obscured by the noise of numerical simulations.

Here, we consider a duplex DNA formed by directed polymers which can
cross each other (see Fig.~\ref{fig:3}). Such models do not show any
melting transition at finite temperatures in $1+1$ dimensions. In the
quantum mechanics analogy, the dimension along which the polymers are
directed plays the role of time: then it is well known that any
short-range potential, no matter how weak, will produce a bound state
for $d\le2$.

Nevertheless, keeping the notation $y=\exp(\beta \epsilon)$ for the
Boltzmann factor associated with base pair interaction, if we
introduce a fugacity $\sigma$ for each bubble formed in the model
between the two DNA strands, it is possible to demonstrate that the
melting transition $y_c(\sigma)$ decreases from $y_c(1)=1$ (i.e.
$(T=\infty)$) for $\sigma=1$ to $y_c(0)=2$ for $\sigma=0$ (i.e. for
the fork model in which bubbles are suppressed.  See Appendix A for
the phase diagram of the fork model.).

\figmodel

In order to generalize the model for three chains (see fig.
\ref{fig:3}) we associate the fugacity $\sigma$ to each bubble opening
between all possible pairs of chains. As regards the base pair
interactions, we consider two options:

\begin{itemize}
\item In Model A we assign a Boltzmann factor $y$ for all two chain
  interactions, but in the case when all three chains come together we
  assign an interaction factor $y^2$ instead of $y^3$.

\item In Model B, we assign a Boltzmann factor $y$ for each
  interaction between chain 1 and chain 2, and between chain 2 and
  chain 3, but we do not consider any interaction between chain 1 and
  chain 3.

\end{itemize}
In both cases, the melting temperatures of the three and two chain
systems coincide at $\sigma=0$, when bubbles are forbidden.  The
existence for a given $\sigma$ of a three chain bound state at a $y<
y_c(\sigma)$, for $0<\sigma<1$, should be a definitive proof of the
Efimov effect for models of triplex DNA.

It is known \cite{Fisher} that in 1+1 dimensions, if a polymer is
confined between two lines or two polymers, there is a repulsive
entropic force of a similar $1/r^2$ potential. In our model B, chain 2
is the only one that can mediate interaction between 1 and 3, because
the latter two chains do not interact, and for chain 2 to do so, it
gets effectively, though not strictly, confined between chains 1 and
3.  In such a situation the steric repulsion and the induced
attractive interaction between 1 and 3 tend to cancel each other\cite{motzkin}. This
weakens the possibility of the Efimov effect in model B but not in the
case of model A.  The nonexistence of the Efimov effect for model B
would therefore be a further proof that our analysis is correctly
taking care of all possible interactions of the models.

The three chain models cannot be solved exactly but they can be
numerically studied with impressive precision through transfer matrix
techniques.  We worked at $\sigma=0.5$. For this value the melting
transitions occurs at $y_m\equiv y_c(0.5)=4/3$.  The existence of the
melting transition can be seen by looking at the rescaled average
distance $\xi_N \equiv r_N/(N^{1/2})$ between the free extrema of an
arbitrary pair of the three chains at a length $N$.  By using standard
finite size scaling techniques, in order to pinpoint the transition
temperature, we computed the intersection between these curves.  The
resulting numbers are presented in Fig.~\ref{fig:4}. In the figure,
besides the data for model A and model B, we also show the data for
the exactly solvable two chains model in order to evaluate the degree
of convergence of the simulations.

\figsimulation 

Beyond any possible numerical uncertainty, model B as we argued
already does not show any Efimov effect while model A exhibits the
Efimov effect.  Although narrow, there exists a temperature interval
for model A in which three chains are bound whereas two are not.

\section{Conclusion}
We have presented a scaling argument to show the possibility of a
three strand DNA bound state at conditions where a duplex DNA would be
in the denatured state.  This is a biological analog of the nuclear or
cold atom Efimov effect.  The scaling argument is further confirmed by
renormalization group and exact numerical results on various model
systems in different dimensions.

We end with a few comments.  To mimic reality, the minimal model
considered here may be supplemented with additional terms like
excluded volume effects whose influence on the phenomenon need
separate analysis.  However if experiments can be done in theta
conditions for the strands, then the excluded volume effects can be
ignored or minimized, and our results would be applicable.  It may be
noted that enzymatic activities are hypothesized to involve local
covalent or hydrogen bonds and contacts. For DNA, one requires
denaturation of strands (melting in some form) locally.  In these lock
and key mechanisms, the issue of nonspecific long range bonds is
generally not considered.  Therefore, the existence of a bound state
involving two otherwise denatured strands of DNA due to the presence
of a third strand, with overall separation much larger than the
hydrogen bond length would have important implications in biological
processes.  Nonetheless, we anticipate new experiments to look for
signatures of our proposed Efimov-DNA.

%%%%%%%%%%%%%%%%%%%%%%%%%%%%%%%%%%%%%%%
\newcommand{\figYph}{%
\begin{figure}[htbp]
   \centering
   \includegraphics[width=0.45\textwidth]{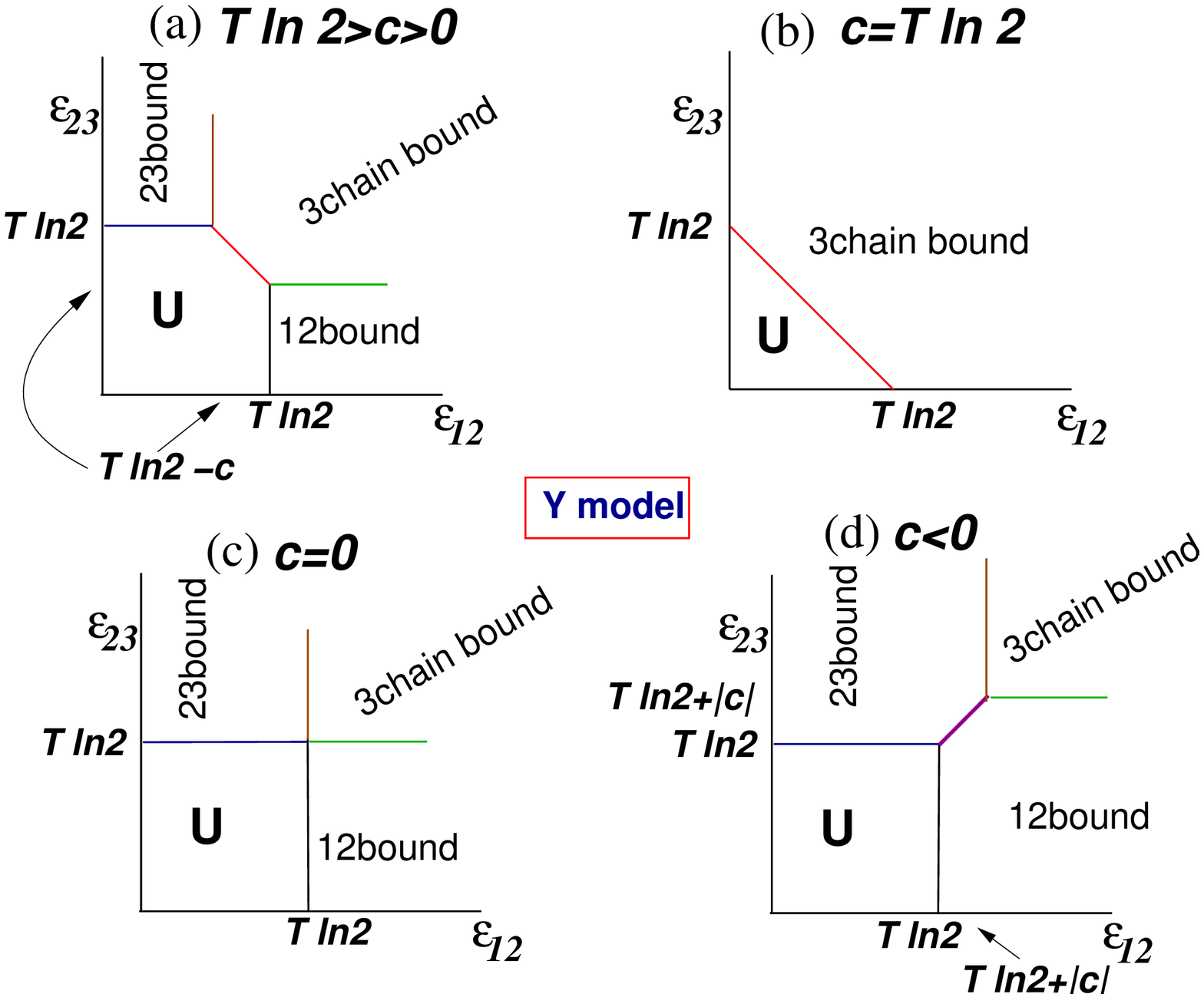}

   \caption{Phase diagrams for 3 chain Y-model. Here $c=\epsilon_{31}$.
     $c<0$ corresponds to repulsive interaction.  ``U" represents
     the unbound state.  }
   \label{fig:yph}
 \end{figure}
}
%%%%%%%%%%%%%%%%%%% Y model part %%%%%%%%%%%%%%%%%%%
\appendix
\section{Phase diagram in absence of loops (fork model)}\label{sec:ymodel}
In order to get the thermodynamic phase diagram, we consider a
simplified model of three chains without bubbles by generalizing the
double chain Y-model studied in the context of melting and unzipping
transition of a duplex\cite{ymodel,ymodel-2}.

Consider 3 chains on a standard square lattice, stretched along one
diagonal direction (directed polymers). The polymers interact with a
pairwise contact interactions $-\epsilon_{i,j}$ between chains $i$ and
$j$. The monomers are indexed from one end $s=0$.  By construction,
the interchain interaction is between monomers with the same index, as
is required for DNA.  There is an additional constraint that an
unbound configuration of two chains cannot be followed by a bound
stretch along the increasing $s$ direction. This avoids loops on the
chains but allows opening of the DNA.  This is the Y-model.  For this
model, every open chain has an entropy $k_B\ln \mu$ per monomer
($\mu=2$ steps per bond) and the same entropy for any bound polymer
(duplex or triplex).  With this entropy and the additive energy, the
free energies per monomer for various possible phases can be written
down.  These are (with $k_B=1$)
\begin{subequations}
\begin{eqnarray}
f_{\rm b} &=& -(\epsilon_{12}+\epsilon_{23}+\epsilon_{31}) -T \ln \mu, 
{\rm( all\  bound),}\ \ \\
f_{\rm u}&=& -3 T \ln \mu,  \quad{\rm(all\  unbound),}\\
f_{ij,k} &=&  -\epsilon_{ij} -2\, T \ln \mu, (ij\ {\rm bound,\ } k\ {\rm free}),
\end{eqnarray}
\end{subequations}
The two chain melting transition is at
$T_{c}^{\{jk\}}=\epsilon_{jk}/(\ln \mu)$.

\figYph

The phase transition lines in this Y-model are all first order lines
whose slopes can be determined by a Clausius-Clapeyron argument.  Let
us take a line of coexistence between two types of phases A and B in a
phase diagram of ``intensive" variables $X_1$ and $X_2$.  Let the
conjugate extensive variables be $\rho_i= {\partial F}/{\partial
  X_i}$, $F$ being the appropriate free energy.  At a given point on
the coexistence curve $\rho_i$ would show a discontinuity, taking
values $\rho_i^{A}$ and $\rho_i^{B}$ in the two phases.  Then the
slope of the line is given by
\begin{equation}
  \label{eq:12}
   \frac{\partial X_2}{\partial X_1} = -
         \frac{\rho_1^A - \rho_1^B}{\rho_2^A-\rho_2^B},
\end{equation}
In our case, $X_1\equiv \beta\epsilon_{12}$,
$X_2\equiv\beta\epsilon_{23}$ and so the derivative of the free energy
gives the associated number of contacts. Consequently, the slope is
related to $\Delta n_{12}/\Delta n_{23}$.  Now in this all-or-none
Y-model, these differences in the fraction of contacts in the two
phases are either 1 or zero.  Therefore, the slopes can in general be
$0$ or $\infty$.  The special case is line $X_1=X_2$.  By symmetry,
here on this line $\Delta n_{12}=\pm\Delta n_{23} $ and the slope
should be $\pm 1$.

If one chain melts away, i.e., a triplex breaks into a duplex and a
free chain, then $T_{b\to12},T_{b\to23}<T_{b\to u}$.  These conditions
follow from the stability of the phases as given by the free-energies.
We therefore have two inequalities: for the $\{12,3\}$-state to occur,
$\epsilon_{12}>\epsilon_{23}+\epsilon_{31}$ and for state $\{23,1\}$
to occur, $\epsilon_{23}>\epsilon_{12}+\epsilon_{31}$.  For easy
display, we represent the three dimensional phase diagram in slices of
the $\epsilon_{12}$-$\epsilon_{23}$ plane for given values of
$c=\epsilon_{13}$ and $T$.  Fig.~\ref{fig:yph} shows four possible
situations for different values of $\epsilon_{31}$ and $T$.  In these
figures, $c=T\ln 2$ is special because of melting of the 1-3 pair.
There is a region in Fig.~\ref{fig:yph}a, where one sees a three chain
bound state in a range of temperatures where none of the pairs would
be bound.  Despite the similarity with the Efimov effect, the bound
state is purely from the pairwise binding and in that sense it is not
a true representative of the effect\cite{allha}.  The figures show
that at or above the melting temperature of the 1-3 pair or if chains
1 and 3 are noninteracting ($c=0$, Fig.~\ref{fig:yph}c), the three
chain bound phase may melt via a duplex or directly to the unbound
state.  Direct melting without an intervening duplex phase is not
possible if $c<0$, (Fig.~\ref{fig:yph}d).

Experimental phase diagrams for triple helix are generally done with
temperature and salt concentration as the two variables.  For a given
concentration, the variation of the temperature would follow a
particular trajectory in the three dimensional thermodynamic phase
space and the possible phases one would see are determined by the
intersection of that curve with the phase boundaries.

%%%%%%%%%%%%%% end of Y-model part %%%%%%%%%%%%%%%

\section{Evidence of a first order transition for triplex}\label{sec:appB}
We present a numerical evidence that there is a discontinuity in the
three chain average energy as shown in Fig.~\ref{fig:1}b.

\figorder

The exact value of $E_n$ at the $n$th generation is computed by using
{\small MATHEMATICA} with infinite precision by iterating the two and
the three chain partition functions and their derivatives.  The length
of the polymers at the $n$th generation is $L_n=2^n$ so that the
thermodynamic limit of energy per monomer $E_n/L_n$ can be obtained by
extrapolation to $1/n\to 0$.  If $C_{n}, Z_n$ and $Q_n$ are the $n$th
generation partition functions for single, double and triple chain
systems, then these obey the recursion relations\cite{smsmb}
\begin{eqnarray}
  \label{eq:4}
  C_{n}&=&bC^{2}_{n-1},\\
  Z_n&=&b(b-1)C^4_{n-1}+bZ^2_{n-1},\\
  Q_n&=&b(b-1)(b-2)C^6_{n-1}\nonumber\\
     &&
     +b(b-1)C^2_{n-1} \sum_{{{i,j=1}\atop{i<j}}}^{3}Z(ij)^2_{n-1}
+bQ^2_{n-1},\label{eq:13}  
\end{eqnarray}
where the arguments of $Z_{n-1}$ in Eq.~(\ref{eq:13}) refer to the two chains
involved.  The initial conditions are 
$$C_0=1, Z_0=y, Q_0=y^3.$$
The relations for the derivatives can be derived from
Eqs.~(\ref{eq:4})-(\ref{eq:13}).

Fig.~\ref{fig:ord} shows the extrapolation in the range of $y=2.323$
to $2.327$ which brackets the transition in the range $(2.324,2.325)$.
The discontinuity survives even on a finer scale in
Fig.~\ref{fig:ord}b, which give $y_0$ in the range $(2.32402,2.32403)$
consistent with the RG result of Fig.~\ref{fig:1}.

%%%%%%%%%%%%%%%%%%%%REFERENCES%%%%%%%%%%%%%%%%%%%%%%%%%%%%%%%%%%%%%%%%%

\end{document}